# Moiré excitons: from programmable quantum emitter arrays to spin-orbit coupled artificial lattices

Hongyi Yu[1], Gui-Bin Liu[2], Jianju Tang[1], Xiaodong Xu[3,4], Wang Yao[1*]

[1] Department of Physics and Center of Theoretical and Computational Physics, University of Hong Kong, Hong Kong, China
[2] School of Physics, Beijing Institute of Technology, Beijing 100081, China
[3] Department of Physics, University of Washington, Seattle, WA 98195, USA
[4] Department of Materials Science and Engineering, University of Washington, Seattle, WA 98195, USA

* Correspondence to: wangyao@hku.hk

**Abstract:** Highly uniform and ordered nanodot arrays are crucial for high performance quantum optoelectronics including new semiconductor lasers and single photon emitters, and for synthesizing artificial lattices of interacting quasiparticles towards quantum information processing and simulation of many-body physics. Van der Waals heterostructures of 2D semiconductors are naturally endowed with an ordered nanoscale landscape, i.e. the moiré pattern that laterally modulates electronic and topographic structures. Here we find these moiré effects realize superstructures of nanodot confinements for long-lived interlayer excitons, which can be either electrically or strain tuned from perfect arrays of quantum emitters to excitonic superlattices with giant spin-orbit coupling (SOC). Besides the wide range tuning of emission wavelength, the electric field can also invert the spin optical selection rule of the emitter arrays. This unprecedented control arises from the gauge structure imprinted on exciton wavefunctions by the moiré, which underlies the SOC when hopping couples nanodots into superlattices. We show that the moiré hosts complex-hopping honeycomb superlattices, where exciton bands feature a Dirac node and two Weyl nodes, connected by spin-momentum locked topological edge modes.

## Introduction

Artificial lattices of interacting particles offer a tunable platform for quantum information processing and quantum simulation of many-body physics, and have been extensively explored for atoms and electrons(*1-3*). Intrigued by the topological states of matter arising from spin-orbit coupling as found in crystalline solids(*4-6*), a significant amount of effort has been devoted to synthesizing spin-orbit coupling in artificial lattices(*2, 3*), which is of great interest for simulating exotic quantum phases such as the topological superfluid.

Van der Waals (vdW) heterostructures of 2D materials provide a new approach towards engineering artificial lattices, where the ubiquitous moiré pattern between different monolayers naturally endows quasiparticles with a nanoscale periodic landscape. The moiré modulation in electronic structures have led to exciting possibilities to engineer topological moiré minibands(*7-11*), and topological insulator superstructures(*12*). In 2D semiconductor vdW heterostrutures, interlayer exciton is a composite quasiparticle that can be most profoundly affected by the moiré, because of the separation of its electron and hole constituents in two layers with interlayer registry varying from local to local (Fig. 1a). Such interlayer excitons have demonstrated bright luminescence, ultralong lifetimes, robust spin-valley polarization, and spin dependent interaction in transition metal dichalcogenides (TMD) heterobilayers(*13-18*), drawing remarkable interest for exploring new spin optoelectronics and high-temperature superfluidity(*19*).

Here we show that the moiré in vdW heterobilayers realizes superlattice potentials in which interlayer excitons have unique spin-dependent complex hopping, leading to giant spin-orbit splitting in the exciton bands. In TMDs heterobilayers, we show the presence of spin-orbit coupled honeycomb superlattices, where the exciton bands feature Dirac and Weyl nodes, and spin-momentum locked edge states dictated by these magnetic monopoles. Hopping in the superlattice potentials can be switched off by a perpendicular electric field, or by strain tuning of moiré periodicity, turning the superlattices into perfect arrays of nanodots that act as uniform quantum emitters. The electric field can also switch the positioning of the quantum emitters in the moiré, hence inverting their spin optical selection rules that are unique imprints of local atomic registries. These properties of moiré excitons point to exciting opportunities towards high performance semiconductor lasers, single photon emitter arrays, entangled photon sources, as well as a platform of versatile tunability for studying exotic quantum phases of matter with imaging possibilities.

**Results**

Fig. 1a-c schematically shows a long period moiré in heterobilayers of $MX_2$ (M=Mo,W; X=Se,S). $MX_2$ monolayers have conduction and valence band edges at K and -K corners of the hexagonal Brillouin zone, where large spin-orbit splitting leads to an effective locking of spin to valley, i.e. the valence band edge at K (-K) has spin up (down) states only. Optically active excitons thus have a pseudospin-1/2 spanned by the spin-valley locked band edges. The various $MoX_2/WX_2$ heterobilayers have the type-II band alignment (Fig. 1d)(*13-16, 18, 20-23*), where electrons (holes) have lower energy in $MoX_2$ ($WX_2$) layer. Excitons thus energetically favor the interlayer configuration with electron and hole constituents in opposite layers.

The locally different interlayer registry in the heterobilayer moiré can be quantified by the in-plane displacement vector $\mathbf{r}_0$ from a metal site in hole layer to a near-neighbor metal site in electron layer. Hopping and vdW interaction between the layers depend sensitively on $\mathbf{r}_0$(*24, 25*). The dependence of $\mathbf{r}_0$ on location $\mathbf{R}$ in the moiré generally results in lateral modulation in local bandgap ($E_g$) and interlayer distance ($d$) (*26, 27*). Interlayer excitons in the moiré thus experience nano-patterned periodic potential:

$$V(\mathbf{R}) = E_g(\mathbf{r}_0(\mathbf{R})) + e\mathcal{E}d(\mathbf{r}_0(\mathbf{R})) - E_b, \qquad (1)$$

where the second term is the Stark shift in a perpendicular electric field $\mathcal{E}$ (*16*). In CVD grown R-type $MoS_2/WSe_2$ heterobilayers, scanning tunneling microscopy/spectroscopy has revealed the laterally modulated local $E_g$ and $d$ in a moiré pattern with period $b = 8.7$ nm (*26*). In Fig. 1c and 1e, the measured local $E_g$ and $d$ values at different $\mathbf{r}_0$ from Ref.(*26*) are shown in comparison with our first principle calculated values for lattice-matched heterobilayer. The exciton binding energy $E_b$ also depends on location $\mathbf{R}$ through $d$, but the dependence is expected to be weak as the variation in $d$ is only a tiny fraction of the in-plane Bohr radius $a_B$ (a few nm).

**Nano-patterned spin optics and programmable quantum emitter arrays**

The description of optical properties of moiré excitons can be facilitated by wavepackets moving adiabatically in the periodic potential. An exciton wavepacket can be constructed using the basis of kinematic momentum eigenstates introduced for lattice-mismatch heterobilayer(*28*) (c.f. supplementary text II), with a real space extension small compared to moiré period $b$, and large compared to monolayer lattice constant $a$. When the wavepacket falls on regions where the local atomic registries preserve the 3-fold rotational ($\hat{C}_3$) symmetry, the wavepacket is an eigenfunction of the $\hat{C}_3$ rotation about its center (c.f. Fig. 2a). Such high symmetry locals are located at *A*, *B* and *C* points of the moiré supercell shown in Fig. 1a, corresponding respectively to the $R_h^h$, $R_h^X$ and $R_h^M$ registries. Here $R_h^\mu$ denotes an R-type stacking with $\mu$-site of electron layer vertically aligned with the hexagon center (*h*) of hole layer. Upon fixing *h* as rotation center, the hole Bloch function at **K** is invariant under $\hat{C}_3$(*29*). The corresponding rotation center in the electron layer is then its *h* center, *X* (chalcogen) site, and *M* (metal) site respectively under $R_h^h$, $R_h^X$ and $R_h^M$ registries, about which the electron Bloch function has distinct $\hat{C}_3$ eigenvalues as shown in Fig. 2a. Hence exciton wavepackets $\mathcal{X}$ at these three locals have the following $\hat{C}_3$ transformations respectively (more details in supplementary text II),

$$\hat{C}_3 \mathcal{X}_{A,s} = e^{-i\frac{2\pi}{3}s} \mathcal{X}_{A,s}, \quad \hat{C}_3 \mathcal{X}_{B,s} = e^{i\frac{2\pi}{3}s} \mathcal{X}_{B,s}, \quad \hat{C}_3 \mathcal{X}_{C,s} = \mathcal{X}_{C,s}. \quad (2)$$

Here $s$ is the spin-valley index: $s = +(-)$ for exciton at **K** (-**K**) valley which has up (down) spin. Photons convertible with excitons must have the same rotational symmetry. Thus Eq. (2) dictates the optical selection rules: the spin up exciton wavepacket at *A* (*B*) couples to $\sigma+$ ($\sigma-$) circularly polarized light only, while light coupling is forbidden at *C*.

The above spin optical selection rules imply the nanoscale patterning of spin optical properties in the moiré. Fig. 2b plots the calculated oscillator strength and polarization of the optical transition dipole as the interlayer exciton wavepacket adiabatically moves in R-type $MoS_2$/$WSe_2$ heterobilayer moiré (c.f. supplementary text I&II), consistent with the symmetry analysis. The optical selection rules cross between the opposite circular polarization at *A* and *B* via elliptical polarization at other locals where the registry no longer has $\hat{C}_3$ symmetry.

The $\hat{C}_3$ symmetry also dictates *A*, *B* and *C* to be the energy extrema of the exciton potential $V(\mathbf{R})$. For R-type $MoS_2$/$WSe_2$ heterobilayers at zero field, the global minima are the *A* points, around which the strong variation in $V(\mathbf{R})$ (~ 100 meV) realizes perfect arrays of nanodot confinement (c.f. Fig. 3a). The strong repulsive interactions(*17*) between interlayer excitons endow these nanodots with two desired functionalities, as single-photon emitters, or as sources of entangled photon pairs, controlled by exciton number loaded per nanodot. As shown in Fig. 3d, the two-exciton configuration, energetically favored in anti-parallel spins because of the repulsive on-site Coulomb exchange ($U_{ex}$), generates a polarization-entangled pair of photons at $E_x$ and $E_x + U_{dd}$ respectively, where $U_{dd}$ is the on-site dipole-dipole interaction. Fig. 3f plots the estimated $U_{ex}$ and $U_{dd}$ in the nanodot confinement of R-type $MoS_2$/$WSe_2$ heterobilayer moiré, both are orders larger compared to the radiative decay rate. Cascaded emissions of multiple excitons thus have well separated spectral resonances.

*B* points are also local minima of $V(\mathbf{R})$, as shown in Fig. 2c. Remarkably, the different local interlayer distances at *A* and *B* (c.f. Fig. 1c) make possible electric field tuning of their energy difference, such that the nanodot confinement of the excitonic quantum emitters can be switched to the *B* points (c.f. Fig. 3a-c). The inversion of $V(A) - V(B)$ happens at a modest field value of $\mathcal{E}_0 = 0.36$ V/nm (0.08 V/nm), taking first principle calculated (STM measured(*26*)) $\delta d$ in R-type $MoS_2$/$WSe_2$ moiré as plotted in Fig. 1c. Along with this electric switching in positions of the quantum emitters, their spin optical selection rules are inverted, and emission wavelength is continuously tuned over a wide range of ~ O(100) meV (c.f. Fig. 3e). Thus, the electric field (interlayer bias) can be used to program the spatial locations, optical selection rule, and emission wavelength of these quantum emitters on demand.

Heterobilayer moiré formed by different TMDs compounds and the H-type stacking feature similar nano-patterned spin optics as the R-type $MoS_2/WSe_2$, while the potential profiles and their field dependence can have quantitative difference. Different compounds combinations and stacking also lead to distinct emission lifetime of the interlayer exciton in nanodot confinement, offering choices on the photon bandwidth in the range of 0.01-1 GHz (see supplementary text II for analysis on the interlayer exciton lifetime). Compared to monolayer TMD emitters(*30-34*), the wider range of choices on the heterobilayer bandgap(*13, 15-17, 20-23, 26*) and significant field tunability further promise applications of these moiré quantum emitters from the visible to the telecom wavelengths.

The lattice mismatch $\delta$ is ~4% between disulfides and diselenides, and ~0.1% between $MoSe_2$ and $WSe_2$ or between $MoS_2$ and $WS_2$. This allows choices on moiré period $b$, over a range up to ~10 nm in M'$S_2$/MSe$_2$ bilayers, and O(100) nm in $MoSe_2/WSe_2$ or $MoS_2/WS_2$ bilayers. A differential strain between the two layers, applicable through a piezoelectric or flexible substrate, further allows *in situ* tuning of $\delta$, which can be magnified into drastic change in the shape and period of the moiré(*12*), which can be exploited for programming the spatial pattern of the emitter array. Exciton hopping between the nanodots is an exponential function of moiré period (Fig. 3f). While for large $b$ or large $\Delta \equiv V(\boldsymbol{A}) - V(\boldsymbol{B})$, the quench of hopping leads to the zero-dimensional spectrum of a uniform quantum emitter ensemble (c.f. Fig. 3g). At $b \leq 10$nm and small $\Delta$, hopping connects the $A$ and $B$ nanodots into an excitonic superlattice (c.f. Fig. 3b).

**Spin-orbit coupled artificial lattices**

Below, we focus on moiré excitons at relative small $b$, and small $\Delta$, where excitons can hop between the $A$ and $B$ dots which form a honeycomb superlattice. The low energy spectrum in such superlattice can be well described by a tight-binding model counting up to third nearest-neighbor (NN) hopping(*35*). Distinct from graphene and other existing artificial honeycomb lattices(*1*), the hopping integrals of interlayer excitons in this superlattice are complex numbers depending on exciton spin *s* and the orientation of the displacement vector between the sites. As illustrated in Fig. 4a, the NN and third NN hopping integrals are (c.f. supplementary text III):

$$t_s(\mathbf{d}_{0/2}) = t_{0/2}, \qquad t_s(\hat{C}_3 \mathbf{d}_{0/2}) = t_{0/2} e^{-i\frac{4\pi}{3}s}, \qquad t_s(\hat{C}_3^2 \mathbf{d}_{0/2}) = t_{0/2} e^{i\frac{4\pi}{3}s}, \qquad (3)$$

where $\mathbf{d}_0 \equiv \frac{b}{\sqrt{3}}\hat{\mathbf{x}}$ ($\mathbf{d}_2 \equiv -\frac{2b}{\sqrt{3}}\hat{\mathbf{x}}$) is a displacement vector from $A$ to a NN (third NN) $B$ site. These spin dependent phases can be derived from the rotational symmetry of the exciton wavefunction at $A$ and $B$ sites, as given in Eq. (2). The second NN hopping integrals are also found to be complex: $t_s(\mathbf{d}_1) = t_s(\hat{C}_3 \mathbf{d}_1) = t_s(\hat{C}_3^2 \mathbf{d}_1) = t_1 e^{i\frac{4\pi}{3}s}$ (c.f. supplementary text III).

Fig. 4b-c plot the exciton dispersion in such a complex-hopping honeycomb lattice with $\Delta = 0$, exhibiting a giant spin-orbit splitting comparable to the bandwidth. The exciton bands feature a Dirac node (4-fold degenerate), and two Weyl nodes (2-fold degenerate) with opposite spin polarization. At a zigzag boundary, a nearly flat edge band connects the two Weyl nodes through the Dirac node at which the spin-polarization of the edge band flips sign. These topological features originate from the spin-dependent hopping phases, which render the Hamiltonian of spin-*s* block a graphene model shifted in the momentum space by $s\mathbf{K}_m$, the wavevector at a moiré-Brillouin-zone (m-BZ) corner. The two spin-up Dirac cones now appear at the m-BZ center $\mathbf{\Gamma}$ and the m-BZ corner $-\mathbf{K}_m \equiv (0, -\frac{4\pi}{3b})$ respectively (c.f. Fig. 4d), while the spin-down Dirac cones are located at $\mathbf{\Gamma}$ point and $\mathbf{K}_m$ corner respectively.

The edge band has its topological origin from these Weyl nodes in the bulk bands. In each spin subspace, the pair of Weyl nodes are linked by an incomplete edge band, which are analog of the surface Fermi arc in 3D Weyl semimetals(*6*) and have been addressed in graphene(*36*). The spin up and down edge bands now appear in conjugate moment space regions and are joined together at $\mathbf{\Gamma}$ point by the coincidence of the spin up and down Weyl nodes that constitute a Dirac node. Observation of these spin-dependent excitonic edge modes can provide evidence of the topological nature.

The 2D Dirac/Weyl nodes will get gapped by a finite $\Delta$, the Dirac mass, tunable by the electric field $\mathcal{E}$ (Fig. 3a-c). Inter-conversion between exciton and photon happens in a momentum space regime known as the light cone (c.f. Fig. 4b), where the energy-momentum conservation can be satisfied. Since the light cone encloses only the Dirac cone at $\mathbf{\Gamma}$ point, the tuning of the Dirac mass from zero to finite by the interlayer bias can drastically change the exciton density of states in the light cone and hence the radiative recombination rate of the moiré excitons.

Moreover, the edge modes connecting the 2D Dirac/Weyl nodes can be separately controlled by a potential localized to the boundary(*37*). Fig. 4e shows an example of such control, where the edge band can be continuously tuned from nearly flat band to the gapless helical states by tuning the magnitude of the onsite energy at the outermost column of A sites (c.f. Fig. 4a). This points to interesting possibilities of experimental manipulation of topological exciton propagation and light coupling on the edge.

These band topological features, together with the observed spin-valley dependent exciton interactions(*17*), further point to a unique Bose-Hubbard system for exploring many-body phenomena with the electric and strain tunability on the hopping and interactions (Fig. 3f).

**Discussion**

Most of the properties discussed here are not limited to TMD heterobilayers, and can be generally expected in other long-period heterobilayer moiré that host interlayer excitons. The patterned optical properties are imprints on the excitons by their local atomic registries, which sensitively affect the interlayer hopping and hence exciton's optical dipole moment. The spin-dependent complex hopping is also a generic feature of interlayer valley exciton in the heterobilayer moiré. Under proper gauge choice for the wavepacket, the hopping matrix element from a site at $\mathbf{R}_1$ to another site at $\mathbf{R}_2$ in the moiré superlattices is of the form:

$$t(\mathbf{R}_1 - \mathbf{R}_2) = e^{i(\mathbf{Q}_e - \mathbf{Q}'_h)(\mathbf{R}_1 - \mathbf{R}_2)} |t(\mathbf{R}_1 - \mathbf{R}_2)|, \qquad (4)$$

where $\mathbf{Q}'_h$ ($\mathbf{Q}_e$) is the wavevector at the hole (electron) valley center, not necessarily high symmetry points. The complex hopping phases are fully determined by $\mathbf{Q}_e - \mathbf{Q}'_h$, the momentum space displacement between the electron and hole valley centers due to the lattice mismatch. In the context of TMDs heterobilayers, the displacement wavevector between the electron and hole **K** valleys correspond to the corners of the moiré-Brillouin-zone, and Eq. (4) reproduces all complex-hopping integrals in the honeycomb superlattices given in Eq. (3) and Fig. 4a. In presence of time-reversal symmetry, an exciton of valley configuration ($\mathbf{Q}'_h, \mathbf{Q}_e$) has a time-reversal counterpart at ($-\mathbf{Q}'_h, -\mathbf{Q}_e$) with opposite spin. These two spin species therefore have opposite complex hopping phases. Such spin-dependent complex hopping in general leads to a large effective spin-orbit splitting in the exciton bands (c.f. TMD examples in Fig. 4b-c).

Both the complex hopping and the nano-patterned spin optics reflect the locally different rotational symmetries of exciton wavefunctions because of the locally different atomic registry in the moiré. These truly unique exciton physics enabled by the moiré pattern do not have counterpart in other systems including the GaAs/AlGaAs quantum wells and individual TMD monolayers. TMD heterobilayer moiré thus provide unprecedented opportunities to explore excitonic quantum emitters and exciton superlattices.

In the typical moiré pattern, the spacing of the nanodots ranges from several to several tens nm, in the regime where a dense array of uniform exciton emitters can be collectively coupled to common optical modes. The moiré quantum emitter array can thus be exploited for a number of applications and studies, from quantum dot laser, entangled-photon laser(*38*), topological photonics(*39*), to the exploration of the seminal Dicke superradiance phenomena(*40*). Individual quantum emitter in the moiré superlattice can also be addressed utilizing the spatially selective excitation of moiré exciton, for example, using the nano-optical antenna-tip(*41*).

It is worth noting that the *intralayer* excitons also experience moiré-patterned potentials(*11*), much shallower though, due to the dependence of local intralayer gaps on the atomic registry (c.f. Fig. 1d-e). A comparison of the potentials and spin optical selection rules for the *inter-* and *intra-layer* excitons in R-type $MoS_2/WSe_2$ heterobilayer moiré can be found in Fig. 2c. Interestingly, the energy minima of $MoS_2$ exciton are at *C*, the optically dark points and energy maxima for interlayer exciton. In contrast, $WSe_2$ exciton has its energy minima at *B*, where the spin-valley selection rules for the *intra-* and *inter-layer* excitons have opposite helicity. These contrasts imply interesting possibilities of dynamic controls for loading the quantum emitter arrays.

## Materials and Methods

The first-principles calculations are performed using the Vienna Ab-initio Simulation Package (*42*) based on plane waves and the projector-augmented wave (PAW) method (*43*). The Perdew-Burke-Ernzerhof (PBE) (*44*) exchange-correlation functional is used for all calculations and the van der Waals interactions are considered in the DFT-D3 (*45*) method. The experimentally measured bulk lattice constants are respectively 3.288 Å for $MoSe_2$ (*46*), 3.160 Å for $MoS_2$ (*46*), and 3.282 Å for $WSe_2$ (*47*). Their average of 3.285 Å (3.221 Å) is used for the lattice-matched $MoSe_2/WSe_2$ ($MoS_2/WSe_2$) heterobilayer. Keeping the in-plane positions fixed, the out-of-plane positions are relaxed for all atoms until the energy difference of successive atom configurations is less than $10^{-6}$ eV. The out-of-plane force on each atom in the relaxed structure is less than 0.003 eV/Å. The cutoff energy of the plane-wave basis is set to 350 eV and the convergence criterion for total energy is $10^{-8}$ eV. A $\Gamma$-centered **k** mesh of $15 \times 15 \times 1$ is used for both the relaxation and normal calculations. The thickness of vacuum layer is greater than 20 Å to avoid impacts from neighboring periodic images. Spin-orbit coupling is taken into account for all calculations except in structure relaxation.

**Acknowledgments:** The work is supported by the Croucher Foundation (Croucher Innovation Award), the Research Grants Council (HKU17302617) and University Grants Committee of Hong Kong (AoE/P-04/08), and the University of Hong Kong (ORA). G.-B.L. was supported by NSFC with Grant No. 11304014 and the China 973 Program with Grant No. 2013CB934500. X.X. was supported by Department of Energy, Basic Energy Sciences, Materials Sciences and Engineering Division (DE-SC0008145 and SC0012509), and the Cottrell Scholar Award.

**Author contributions:** W.Y. conceived and designed the research. H.Y. performed the calculations and analysis, with input from W.Y., X.X. and J.T.. G.B.L provided support with first-principle calculations. W.Y., H.Y. and X.X. wrote the manuscript.

**Competing interests:** The authors declare that they have no competing interests.

**Figures**

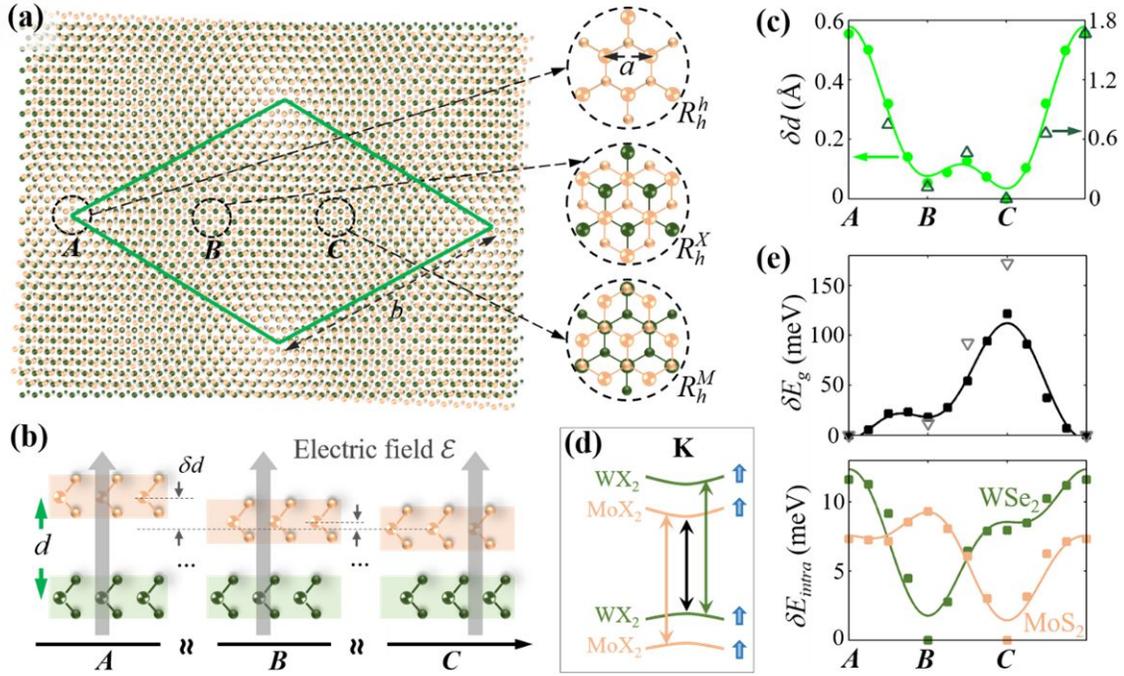

**Figure 1 Moiré modulated local energy gaps and topographic height in heterobilayer**. **a**, Long-period moiré pattern in an $MoX_2/WX_2$ heterobilayer. Green diamond is a supercell. Insets are zoom-in of three locals, where atomic registries resemble lattice-matched bilayers of different R-type stacking. **b** and **c,** Dependence of interlayer distance $d$ on the atomic registries. In **c**, dots are our first principle calculations for $MoS_2/WSe_2$ heterobilayer, and triangles are STM measured variation of local $d$ values in a $b$=8.7 nm $MoS_2/WSe_2$ moiré in Ref. [26]. The variation in $d$ then leads to laterally modulated interlayer bias ($\propto d$) in a uniform perpendicular electric field, as **b** illustrates. **d**, Schematic of relevant heterobilayer bands at K valley, predominantly localized in either $MX_2$ or $WX_2$ layer. **e**, Upper: variation of the local bandgap $E_g$ (black arrow in **d**) in the $MoS_2/WSe_2$ moiré. Lower: variation of local *intralayer* gaps (c.f. arrows of the same color in **d**). In **c** and **e**, the horizontal axis corresponds to the long diagonal of moiré supercell, and the vertical axis plot the differences of the quantities from their minimal values. The curves are fitting of the data points using Eq. (S2-S3) in supplementary text I.

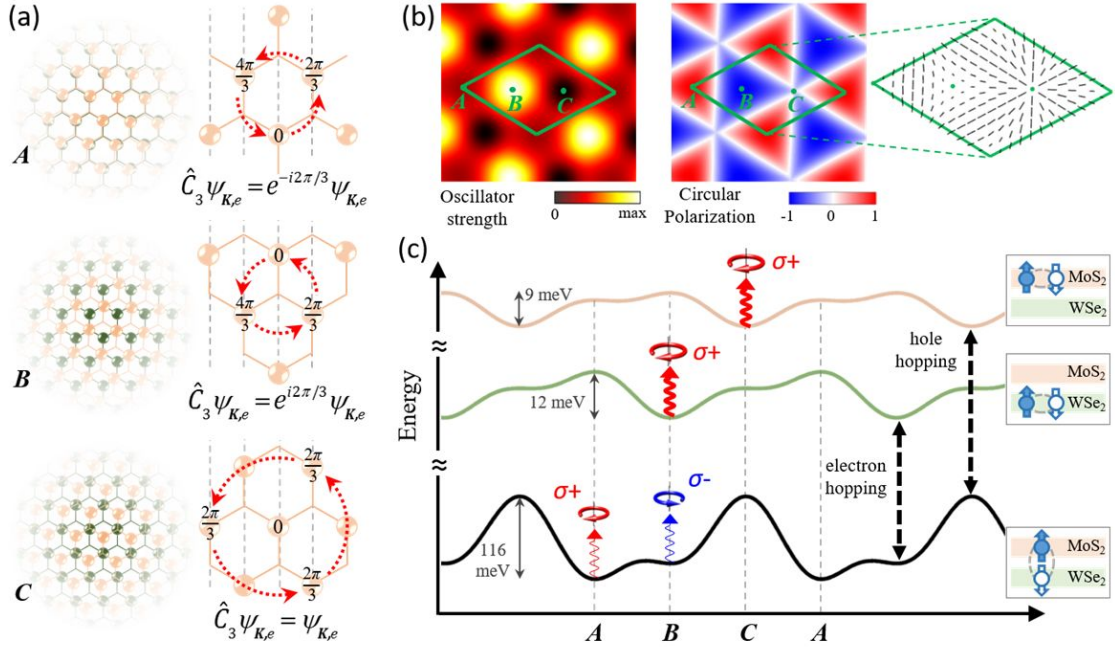

**Figure 2 Nano-patterned spin optics of moiré excitons. a,** Left: Exciton wavepackets at the locals with $R_h^h$, $R_h^X$ and $R_h^M$ registries respectively (c.f. Fig. 1a). Right: Corresponding $\hat{C}_3$ transformation of electron Bloch function $\psi_{K,e}$, when rotation center is fixed at a hexagon center in hole layer. Grey dashed lines denote planes of constant phases in the envelope part of $\psi_{K,e}$, and red arrows denote the phase change by $\hat{C}_3$. **b,** Left: Oscillator strength of interlayer exciton. Right: Optical selection rule for spin-up interlayer exciton (at **K**-valley). The distinct $\hat{C}_3$ eigenvalues as shown in a dictate the interlayer exciton emission to be circularly polarized at *A* and *B* with opposite helicity, and forbidden at *C*. At other locals in the moiré the emission is elliptically polarized (c.f. inset, where ticks denote major axis of polarization with length proportional to ellipticity). **c,** Contrasted potential landscapes for *intralayer* and *interlayer* excitons, with the optical selection rule for the spin-up species shown at the energy minima. Transitions between inter- and intra-layer excitons (i.e. via electron/hole hopping) can be induced by mid-infared light with out-of-plane polarization.

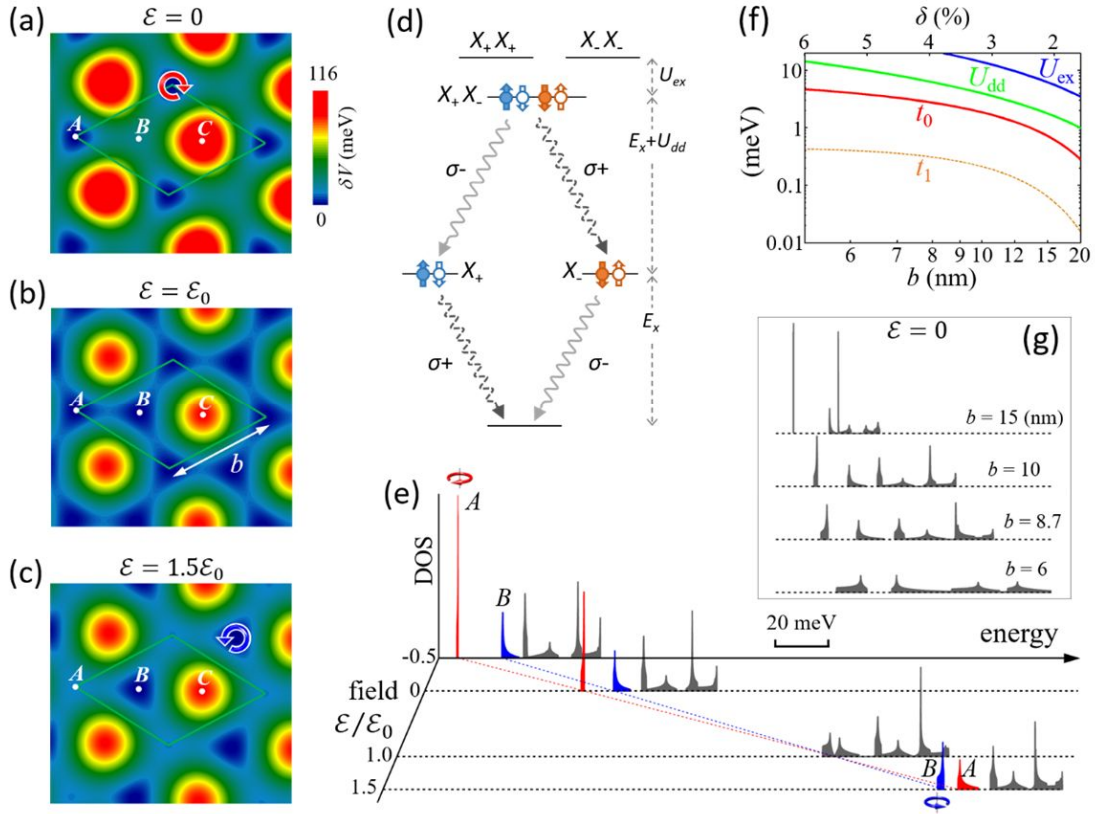

**Figure 3 Electrically and strain tunable quantum emitter arrays. a-c**, Tuning of excitonic potential by perpendicular electric field ($\mathcal{E}$) in R-type MoS$_2$/WSe$_2$. At zero field, nanodot confinements are at **A** points, realizing periodic array of excitonic quantum emitters, which are switched to **B** points at moderate field (c.f. text). **d**, Spin optical selection rule of quantum emitter at **A**. When loaded with two excitons, the cascaded emission generates a polarization-entangled photon pair. The optical selection rule is inverted when the quantum emitter is shifted to **B** (c.f. **a**, **c**). **e**, Electric field tuning of exciton density of states (DOS) in R-type MoS$_2$/WSe$_2$ moiré with $b$=10 nm. The field dependence of $V(\boldsymbol{A})$ and $V(\boldsymbol{B})$ are denoted by the doted blue and red lines on the field-energy plane. Colors of two lowest energy peaks distinguish their different orbital composition at **A** and **B** points in moiré. **f**, Exciton hopping integral between nearest-neighbor **A** and **B** dots in **b** ($t_0$), between nearest-neighbor **A** dots ($t_1$), and onsite exciton dipole-dipole ($U_{dd}$) and exchange ($U_{ex}$) interactions as functions of the moiré period $b$. See supplementary text IV&V. The top horizontal axis is the corresponding lattice mismatch $\delta$ for rotationally aligned bilayer. **g**, Exciton DOS at different $b$ at zero electric field. The 20 meV scale bar applies for the energy axis in **e** and **g**.

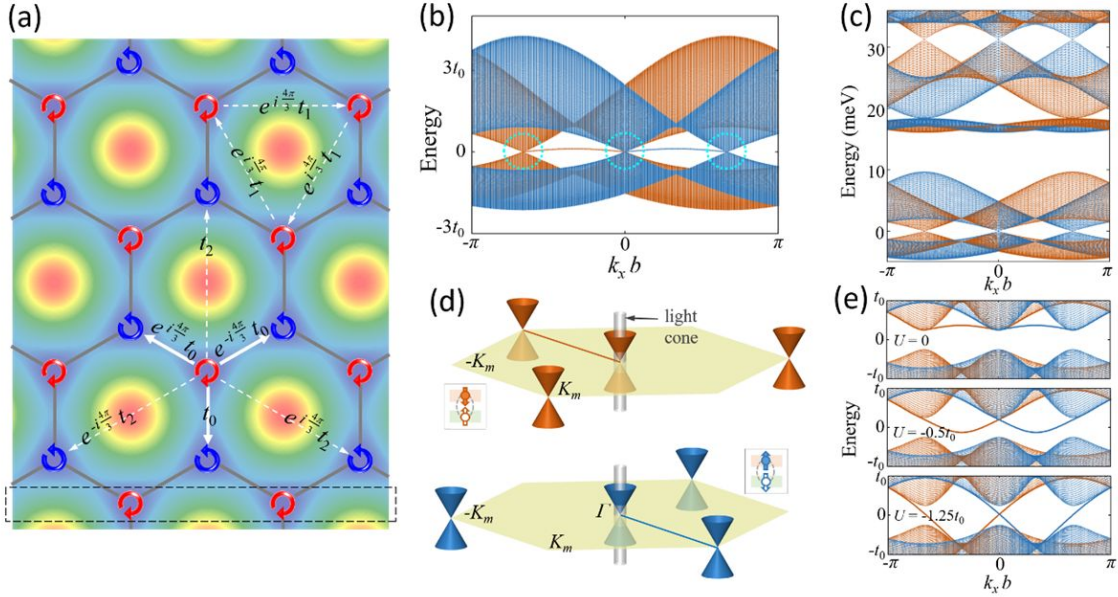

**Figure 4 Spin-orbit coupled honeycomb lattices and Weyl nodes. a**, Opposite photon emission polarization at *A* and *B* sites, and complex hopping matrix elements, for spin up exciton. **b**, Exciton spectrum at $V(A) = V(B)$ and moiré period $b = 10$ nm, from the tight-binding model with third nearest-neighbor hopping. $t_0 = 2.11$ meV, $t_1 = 0.25$ meV, and $t_2 = 0.14$ meV. The bands feature a Dirac node and two Weyl nodes (highlighted by dotted circles). These magnetic monopoles are linked by an edge mode at zigzag boundary, with spin polarization reversal at the Dirac node. Brown (blue) color denotes spin down (up) exciton. **c**, Exact exciton spectrum in this superlattice potential (c.f. supplementary text IV). Dirac/Weyl nodes are also seen in higher energy bands. **d**, Schematic of the Dirac cones for spin up and down excitons in the moiré-Brillouin-zone, and edge modes at a zigzag boundary. Exciton-photon interconversion can directly happen within the shown light cone. **e**, The Dirac and Weyl nodes are gapped by a finite *A-B* site energy difference $\Delta=0.5t_0$, whereas the edge band dispersion is tuned by changing the onsite energy of the dots on the zigzag boundary (enclosed by the dashed box in **a**) by the amount $U$.